\documentclass[twoside]{LCWS11}
\usepackage[latin1]{inputenc}
\usepackage[dvips]{graphicx,epsfig,color}
\usepackage{wrapfig,rotating}
\usepackage{amssymb,amsmath,array}
\usepackage{cite}
\begin{document}
\title{
Iowa Particle Flow Algorithm} 
\author{G. Halladjian, U. Mallik and R. Zaidan
\vspace{.3cm}\\
University of Iowa, Iowa City, Iowa 52242, USA
}

\maketitle

\begin{abstract}
The particle detectors at the future linear colliders, like ILD and SiD, use Particle Flow Algorithms (PFA)s to reach higher jet energy resolutions than the classical pure calorimetry. During the past few years, the University of Iowa group developed the Iowa PFA. This algorithm has been used to benchmark the performance of the SiD detector for the Letter Of Intent of 2009 \cite{sid_loi,matt_charles}. Recently, new strategies and techniques are included in the different parts of this algorithm in order to increase its performance. The latest improvements and results of the Iowa PFA will be discussed.
\end{abstract}

\section{Particle Flow Algorithms}


A Particle Flow Algorithm (PFA) is an algorithm that aims to reduce jet energy resolution by reconstructing the 4-momentum of each stable particle in a jet separately and using the appropriate subdetector to measure the energy depending on the type of the particle. In practice, a PFA attempts to separate electromagnetic energy from hadronic energy, where the hadronic energy is further separated into charged and neutral energy. The electromagnetic calorimeter (ECal) is used for the measurement of the electromagnetic energy, while the hadronic calorimeter (HCal) is used with the ECal for the measurement of neutral hadronic energy. Finally, the inner tracker is used for the measurement of the charged energy.\\
A well performing PFA relies on high precision tracking system and a high granularity calorimeters in order to correctly assign each hit to its corresponding particle. In practice, inevitable confusion has the effect of smearing the jet energy resolution obtained with a PFA. This resolution is typically given by:
\begin{equation}
\sigma_{E} = \sigma_{em} \oplus \sigma_{nh} \oplus \sigma_{ch} \oplus \sigma_{confusion}
\end{equation} 
where $\sigma_{em}$, $\sigma_{nh}$ and $\sigma_{ch}$ are the energy resolutions of electromagnetic particles, neutral hadrons and charged hadrons respectively, and $\sigma_{confusion}$ is an additional term that represents the effects of confusion. $\sigma_{em}$ and $\sigma_{nh}$ are limited by the performance of the calorimeter systems, while $\sigma_{ch}$ is considered to be negligible since the inner tracker provides much better precision than what calorimeters can achieve. The remaining term $\sigma_{confusion}$ caracterises the performance of the PFA for the specific detector.

\section{The SiD Iowa PFA}

The PFA that we present in this document, was developed by the University of Iowa and applied in the context of the SiD detector at the ILC. The best jet energy resolution that can be achieved in the ideal case neglecting the confusion is roughly\footnote{The energy ($E$) here is measured in GeV.} $20\%/\sqrt{E}$ \cite{ron_cassell}.\\
The Iowa PFA consists of two main steps. The first step is a setup stage where the electromagnetic energy is separated, a first clustering is applied on the hadronic energy to identify large and small structures, and a matching is performed between inner detector tracks and calorimeter hits. The second step is the shower reconstruction where the small structures are linked to each other to build large showers.

\subsection{Setup of the Iowa PFA}\label{sec_setup}

The first step of the Iowa PFA consists of reconstructing and identifying electromagnetic clusters in the ECal which are then categorized into electrons or photons depending on whether a track can be matched to the cluster. The calorimeter hits and the tracks used in this stage are removed and a muon reconstruction and identification step is applied.\\
After separating out photons, electrons and muons, the remaining hits and tracks are used for hadronic particle reconstruction. A pre-shower minimum ionizing particle (MIP) finding algorithm follows: tracks in the inner tracker are extrapolated to the calorimeters and matched to isolated hits.\\
The remaining hits are used by a directed tree clustering algorithm (DTree) which identifies large structures in the calorimeter by grouping hits around local maxima in hit density. The DTree is followed by several algorithms to find substructures inside the large clusters and classify them as:
\begin{itemize}
\item MIPs: stubs of isolated hits representing minimum ionizing particles.
\item clumps: clusters of high local density.
\item blocks: essentially large and dense structures found by the DTree algorithm that couldn't be broken into substructures.
\item leftover hits: low density hits that are not used in any structure. These hits are left out of the shower building process, and their energy is shared using 
appropriate weighting technique, among the MIPs, clumps and blocks.
\end{itemize}
At this stage, a second attempt to connect a subcluster to tracks that couldn't be matched to a pre-shower MIP is made sequentially to MIPs, clumps, blocks, and leftover hits.\\
Finally, an attempt to recover hadronic energy that might have been identified as photons is made based on the full information available so far. This step is referred to in the following as {\it photon veto}, in the sense that the decision taken by the photon ID can be vetoed and the cluster in question put back to the hadron energy pool.

\subsection{Shower reconstruction}

The shower reconstruction step attempts to link together MIPs, clumps and blocks that belong to the same shower development. The linking in the baseline algorithm \cite{matt_charles} is based on a score definition that combines a likelihood, several {\it ad hoc} penalty factors and a cone algorithm which gives high scores to subclusters lying along the shower axis. This reconstruction is constrained with a rather tight energy/momentum balance to avoid unphysical expansion of the showers.\\
To avoid mistakes in cases where a high momentum track steals energy from low momentum tracks, the reconstruction is performed in increasing order of track momentum.

\section{Improvements made to the Iowa PFA}

In this section, we discuss modifications applied to the baseline algorithm.

\subsection{Clump finding algorithm}

\begin{wrapfigure}{r}{0.5\columnwidth}
\centerline{\includegraphics[width=0.45\columnwidth]{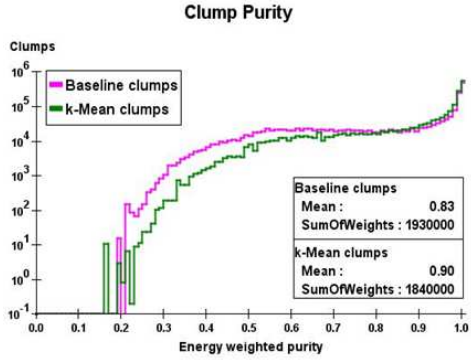}}
\caption{The purity of the clumps reconstructed by 2 methods: NN and $k$-mean complemented by NN.}\label{fig_purity}
\end{wrapfigure}

The baseline clump finding algorithm consists of running a nearest neighbor (NN) clustering algorithm on a selection of hits with high energy density. This algorithm was found to be suboptimal in the enviroment with high overlap, especially in the HCal where the granularity is lower and the analog hit energy measurement is absent. This clump finding strategy has been replaced by the $k$-mean clustering algorithm (described below), complemented by the old NN algorithm to enhance efficiency.\\
The $k$-mean algorithm consist of two steps:
\begin{itemize}
\item A {\it core finding} step where $k$ initial cluster cores are defined as local density maxima.
\item A {\it clustering} step where each hit is assigned to the {\it closest} seed where the metric used is the geometrical distance between the hit and the nearest hit in the core.
\end{itemize}
Figure \ref{fig_purity} shows the distributions of the clump purity with the baseline and the new clump finding algorithms, which was improved from $83\%$ to $90\%$. The purity in this context is the highest fraction of energy used in the cluster that originate from the same particle.\\
\begin{wrapfigure}{r}{0.5\columnwidth}
\centerline{\includegraphics[width=0.45\columnwidth]{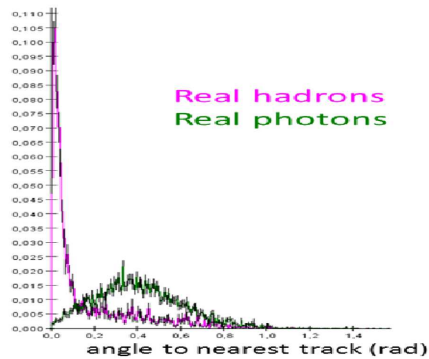}}
\caption{The distribution of the angle between a photon and the nearest track for real hadrons and real photons.}\label{fig_photonVeto}
\end{wrapfigure}
The $k$-mean algorithm is able to break the big structures to smaller pieces with increased purity. This increase in purity is achieved at the cost of small decrease of the fraction of the total event energy which goes to clumps, from $47.5\%$ to $45\%$. This lost energy showes up as an increase in the leftover hits.

\subsection{Photon veto}

In some cases, energy deposited by secondary $\pi^0$ mesons originating from hadronic showers, can be incorrectly identified as primary photons. In the baseline algorithm, an identified photon is vetoed if pre-shower MIP reconstruction attempts to use hits already attributed to the photon. This criterion was found to be too aggressive to real primary photons, since it is frequent that the pre-shower MIP finding uses hits from the {\it halo} of the photon. In fact, $40\%$ of the real photons were being vetoed by this criterion.\\
This strategy has been revisited and a new criterion was defined: an identified photon is treated as a hadronic clump if it is within an angle of $3$ degrees of a reconstructed track.\\
Figure \ref{fig_photonVeto} shows the distribution of the angle between an identified photon and the nearest reconstructed track, for real primary photons and secondary hadronic clusters identified as photons.

\subsection{Track-seed matching}

\begin{figure}[!ht]
\begin{minipage}[t]{.5\textwidth}
\includegraphics[width=1\textwidth]{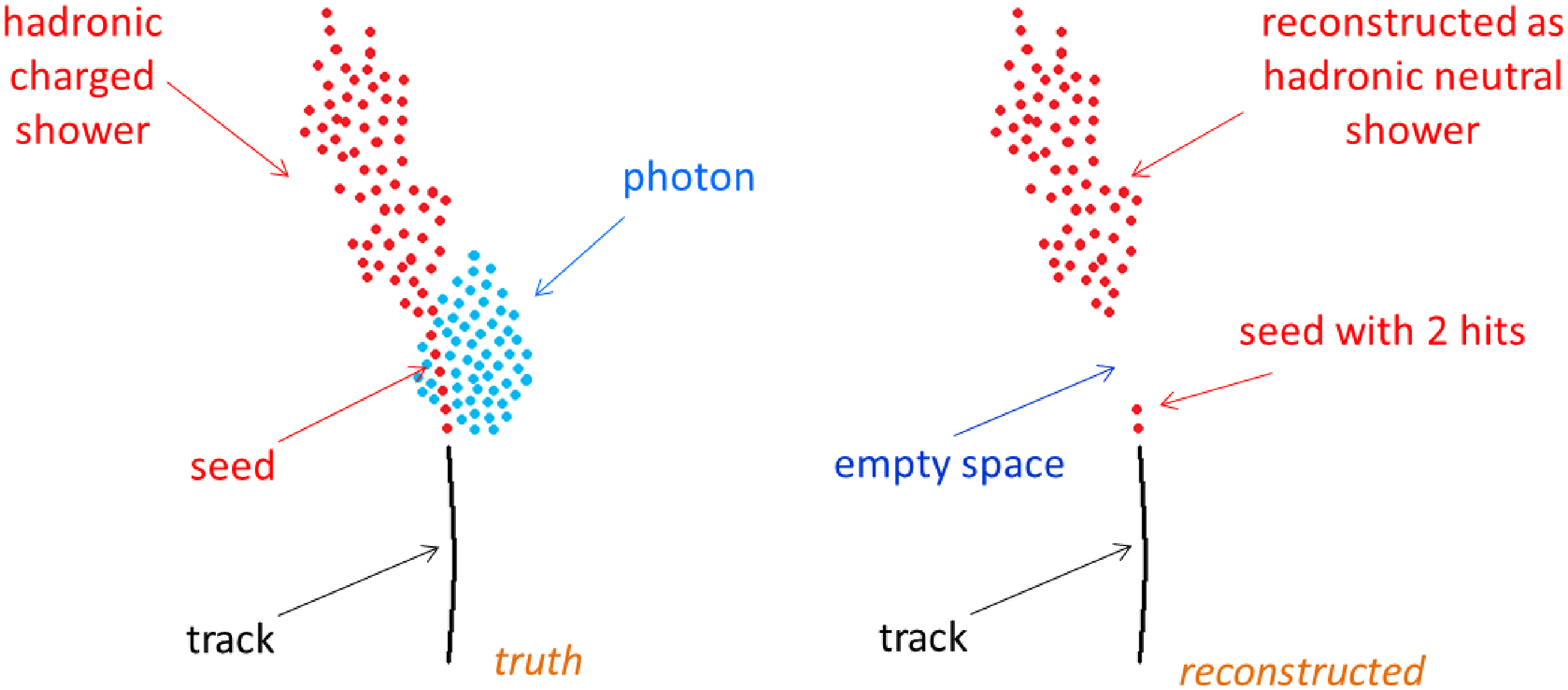}
\caption{An illustration of the first case of improvement of track seed matching.}
\label{fig_trackSeedMatching1}
\end{minipage}
\hfill
\begin{minipage}[t]{.4\textwidth}
\includegraphics[width=1\textwidth]{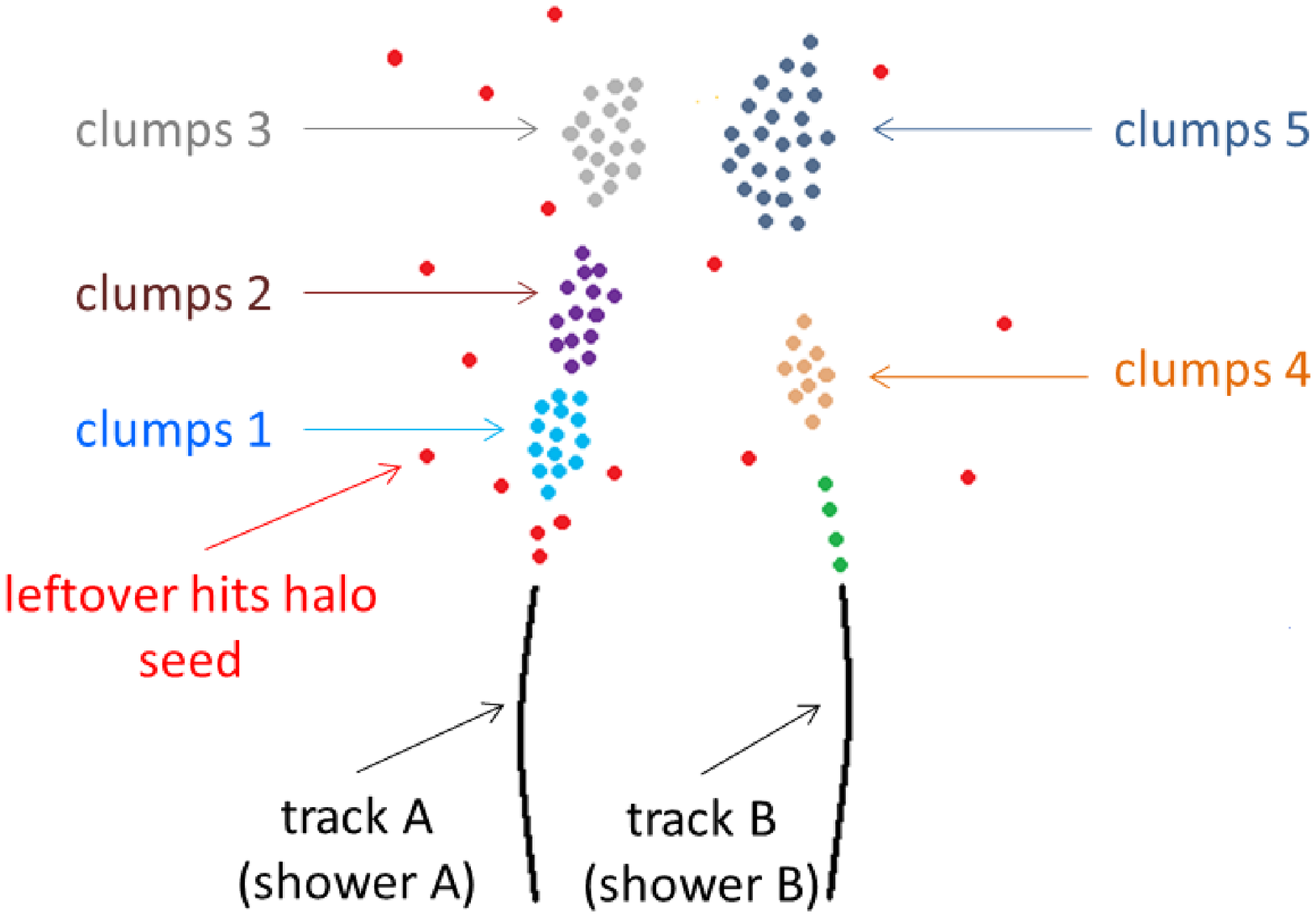}
\caption{An illustration of the second case of improvement of track seed matching.}
\label{fig_trackSeedMatching2}
\end{minipage}
\end{figure}

Improvements were made in two types of situations. For almost $8\%$ of the tracks with a close-by photon, a large number of hits from the track is ``claimed'' by the photon. These hits, when removed as part of the photon, leaves a gap prohibiting the propagation of the shower further (Figure \ref{fig_trackSeedMatching1}).\\
The second improvement applies to about $7\%$ of the tracks, when a track is matched to a halo of leftover hits after having failed to be matched to a MIP, a clump or a block. The halo in question occupies a large volume and can cause confusion during the linking process as it is shown in Figure \ref{fig_trackSeedMatching2}.\\
Both of these cases, once identified, are solved by using a helix extrapolation of the track into the calorimeter independent of the depth. A new matching is then attempted sequentially to MIPs then clumps. Of these $15\%$ of instances, almost $80\%$ are fixed by this procedure.

\subsection{Link scoring}
 
\paragraph{The scoring:\\}

\begin{wrapfigure}{r}{0.5\columnwidth}
\centerline{\includegraphics[width=0.35\columnwidth]{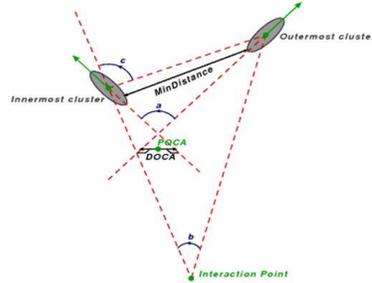}}
\caption{The variables used for the link scoring.}\label{fig_proceedingPlot5}
\end{wrapfigure}

\begin{figure}[!ht]
\begin{minipage}[t]{.45\textwidth}
\centerline{\includegraphics[width=\textwidth]{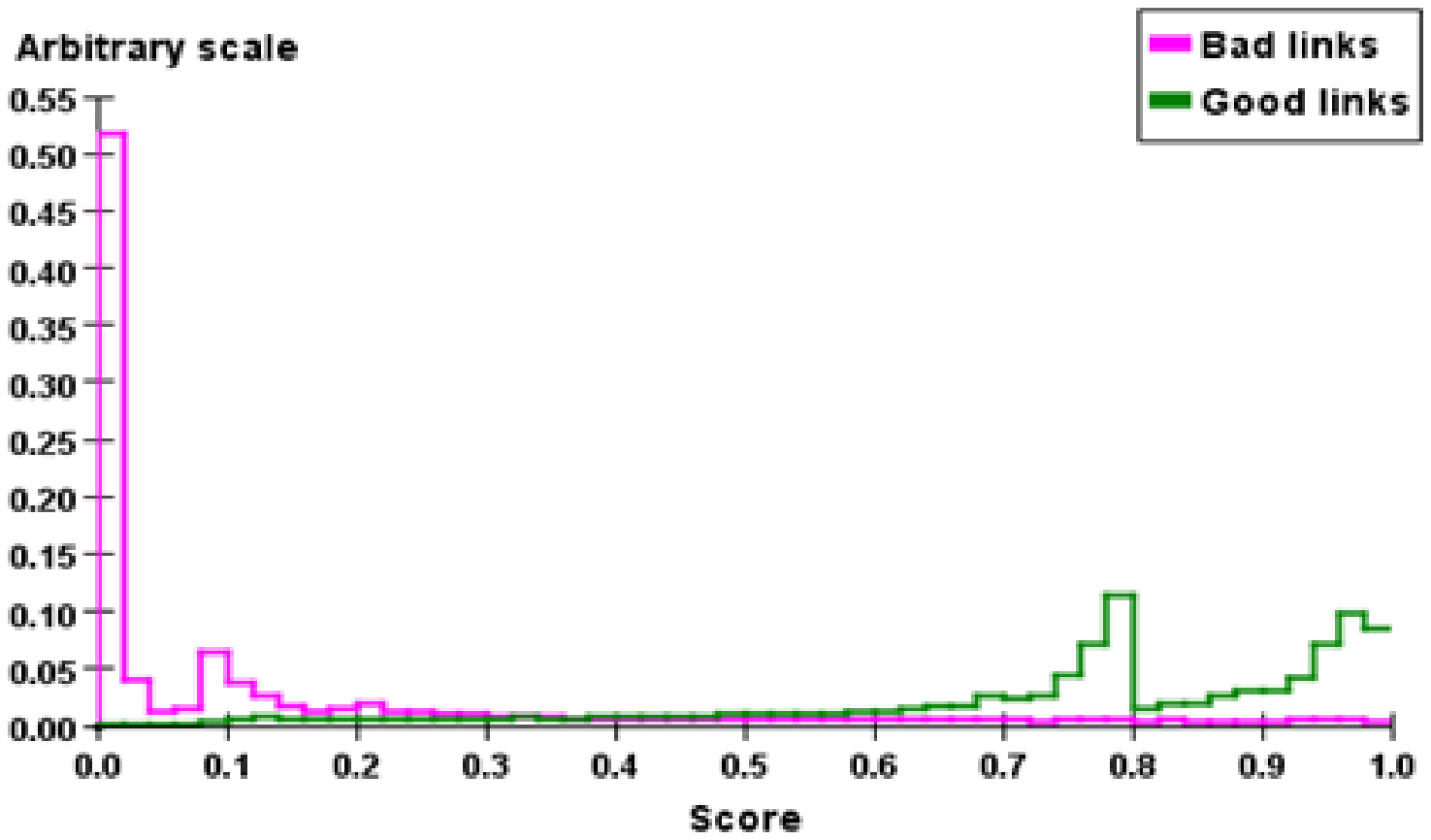}}
\caption{The likelihood distribution for good and bad links obtained with the baseline scoring.}
\label{fig_proceedingPlot6}
\end{minipage}
\hfill
\begin{minipage}[t]{.45\textwidth}
\centerline{\includegraphics[width=\textwidth]{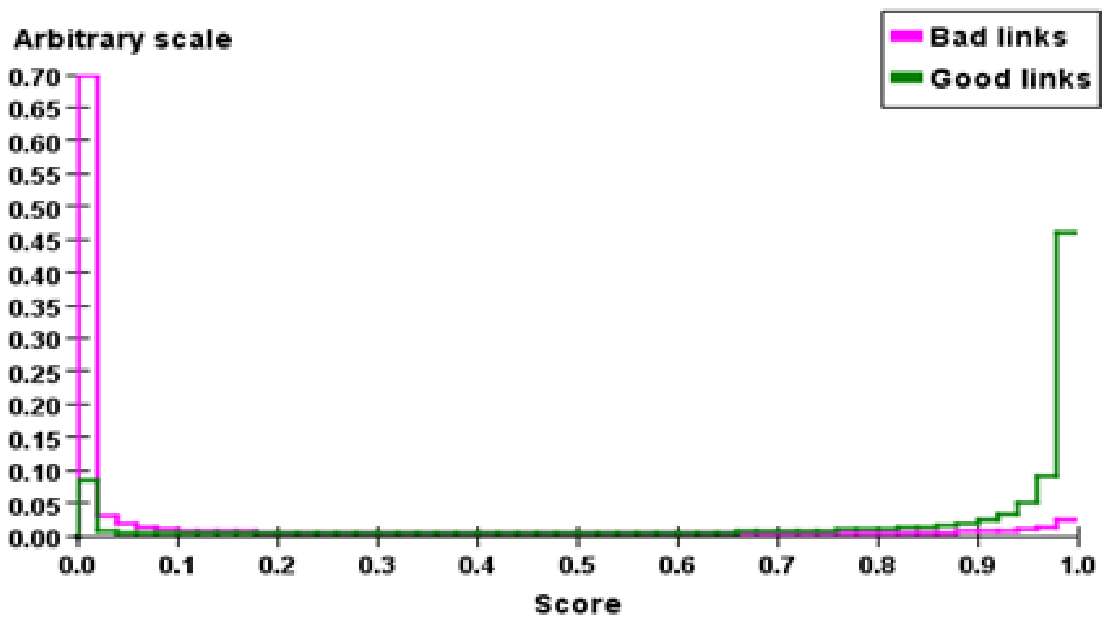}}
\caption{The likelihood distribution for good and bad links obtained with the improved scoring.}
\label{fig_proceedingPlot7}
\end{minipage}
\end{figure}

The scoring used for the baseline PFA is only partially based on a likelihood. Several {\it ad hoc} penalty factors based on angular distance and proximity of the two clusters to be linked were introduced on top of the likelihood. The scoring process is now much more elaborate with addition of new variables and introduction of correlations among the variables with all the {\it ad hoc} conditions removed. These new variables, shown in Figure \ref{fig_proceedingPlot5}, are:
\begin{itemize}
\item the angle between the directions of the two clusters (angle $a$).
\item the kink angle as seen from the interaction point (angle $c$).
\end{itemize}
The correlations between the different variables used for the likelihood are taken into account by defining two dimensional probability density functions (PDF)s when needed.\\
Variations in the properties of different sections of the detector are now taken into account by defining different probability density functions (PDF). 
\paragraph{The training:\\}
To produce the PDFs for the likelihood, simulated Monte Carlo information is used to determine whether a link should be used as a {\it good} or a {\it bad} link. In the baseline PFA, the link between two subclusters originating from the same primary particle was defined as a good link, even if the two subclusters were produced by two different secondary particles in the shower developpement. This definition has the disadvantage of deluting the discrimination of the likelihood. The definition is changed to only treat as good links, immediate and direct links between subclustes originating from the same primary particle.
\paragraph{The improvement:\\}
Figures \ref{fig_proceedingPlot6} and \ref{fig_proceedingPlot7} show the likelihood distributions for good and bad links as obtained with the baseline scoring and with the new scoring respectively.

\section{Results}

\begin{figure}[!ht]
\begin{minipage}[t]{.45\textwidth}
\centerline{\includegraphics[width=\textwidth]{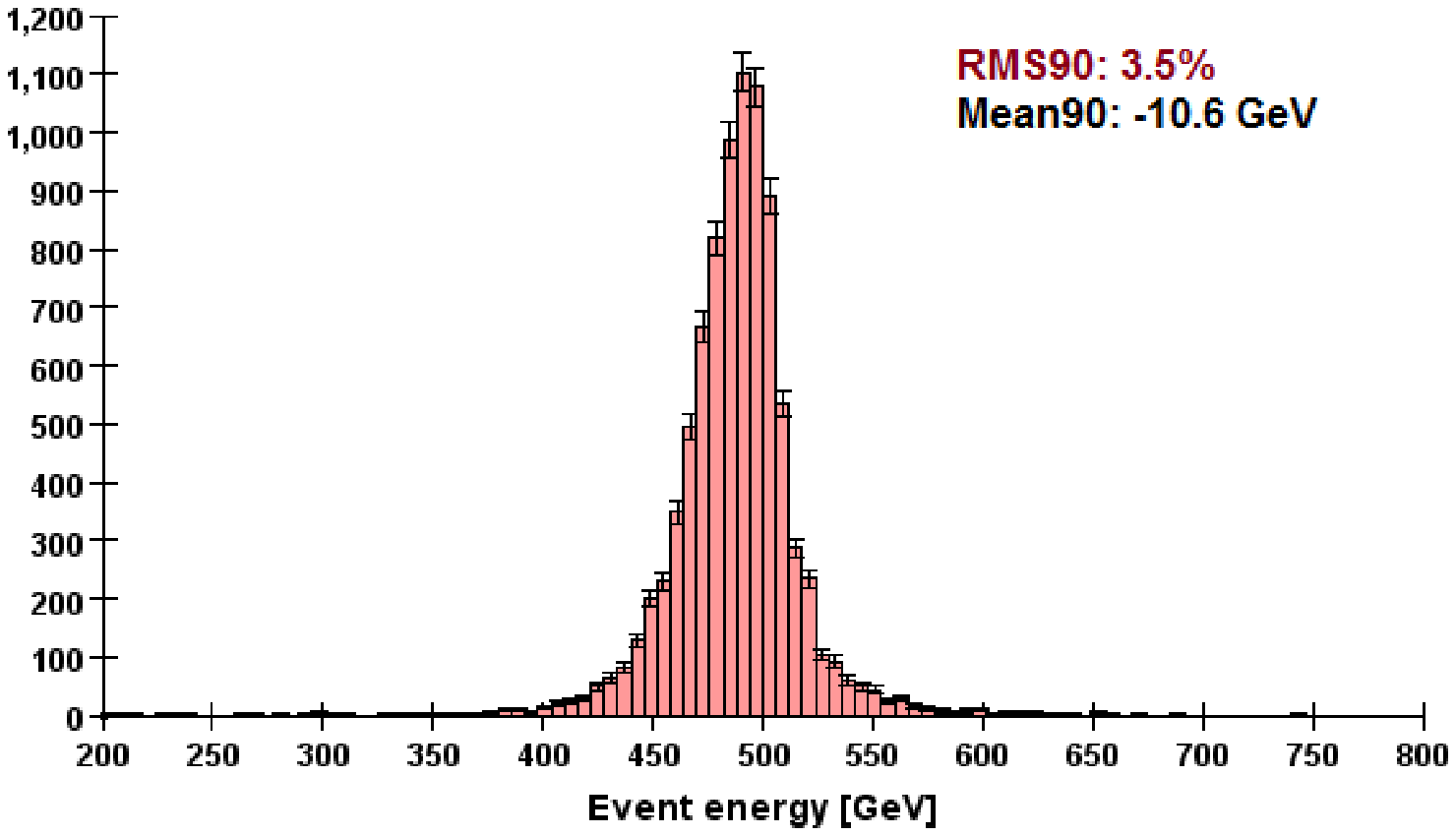}}
\caption{The energy residual distribution obtained with the baseline PFA.}
\label{fig_proceedingPlot8}
\end{minipage}
\hfill
\begin{minipage}[t]{.45\textwidth}
\centerline{\includegraphics[width=\textwidth]{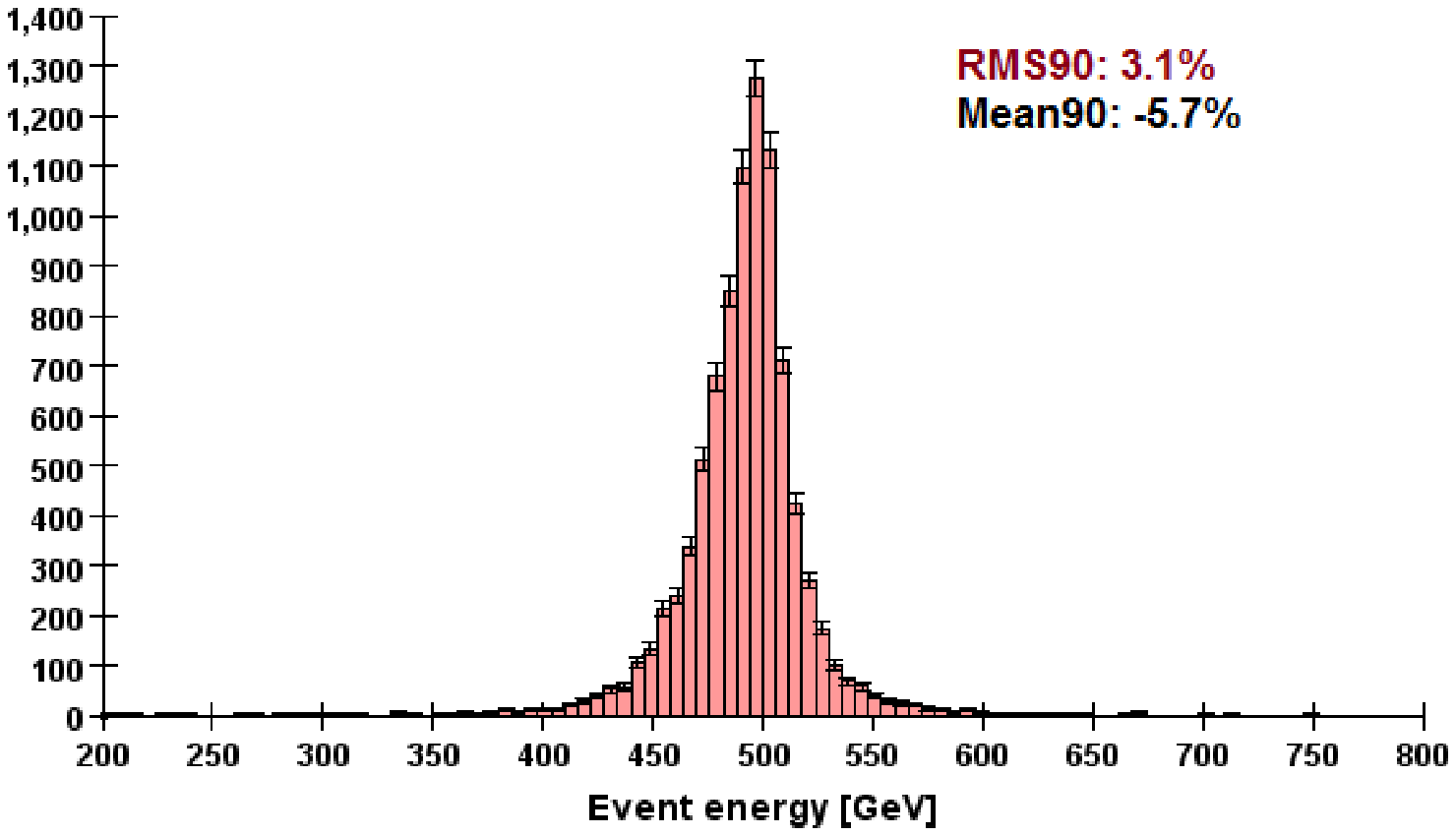}}
\caption{The energy residual distribution obtained with the improved PFA.}
\label{fig_proceedingPlot9}
\end{minipage}
\end{figure}

The energy resolution obtained with simulated $e^+e^-\to q\bar q$ event at 500 GeV center of mass energy, with the modifications mentioned above, improves to $3.1\%$ compared to $3.5\%$ with the baseline PFA. Figures \ref{fig_proceedingPlot8} and \ref{fig_proceedingPlot9} show the event energy residual distributions obtained with the baseline PFA and with the described improvements respectively.

\begin{figure}[!ht]
\begin{minipage}[t]{.45\textwidth}
\centerline{\includegraphics[width=\textwidth]{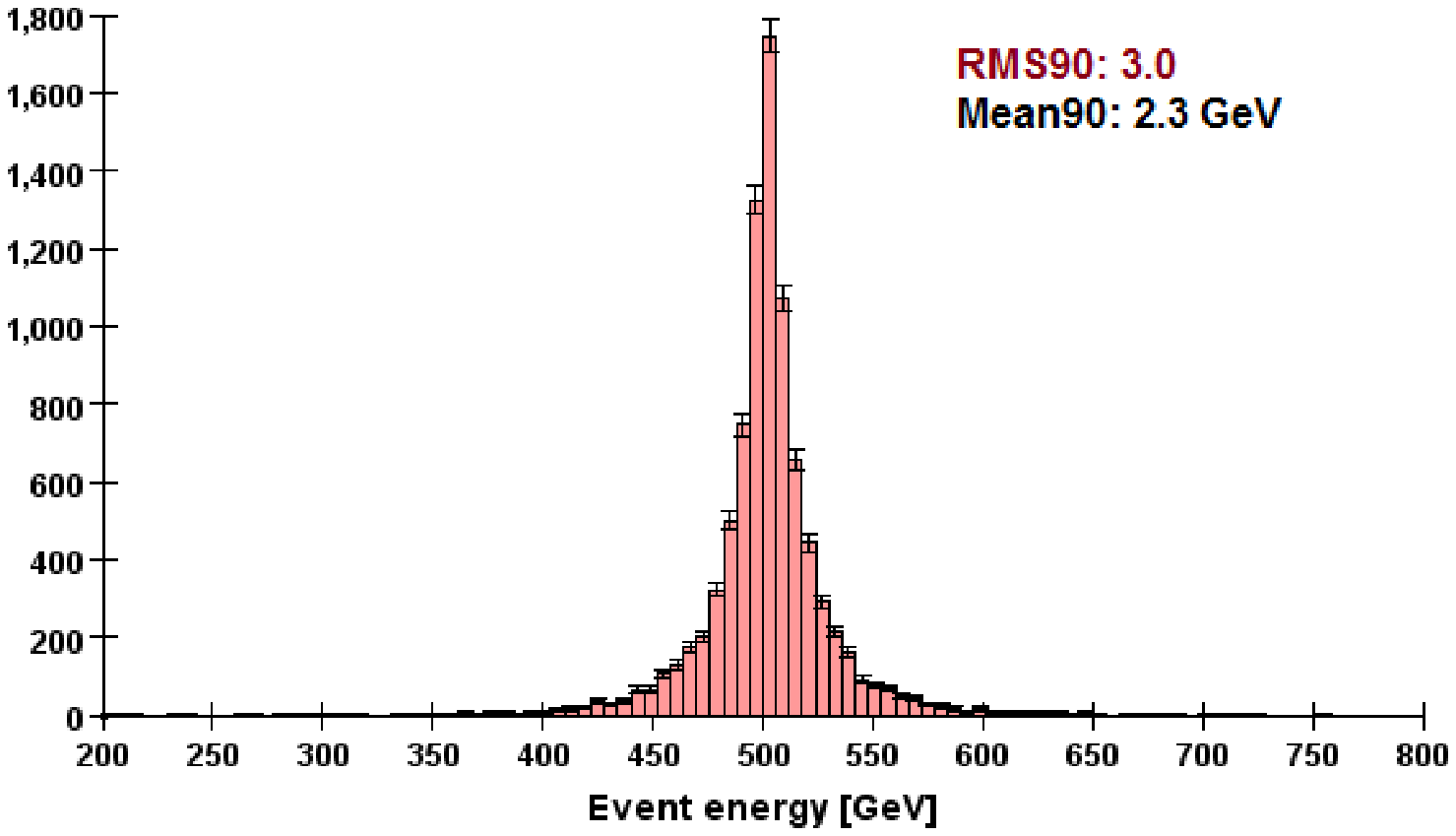}}
\caption{The energy residual distribution obtained with perfect shower building.}
\label{fig_proceedingPlot10}
\end{minipage}
\hfill
\begin{minipage}[t]{.45\textwidth}
\centerline{\includegraphics[width=\textwidth]{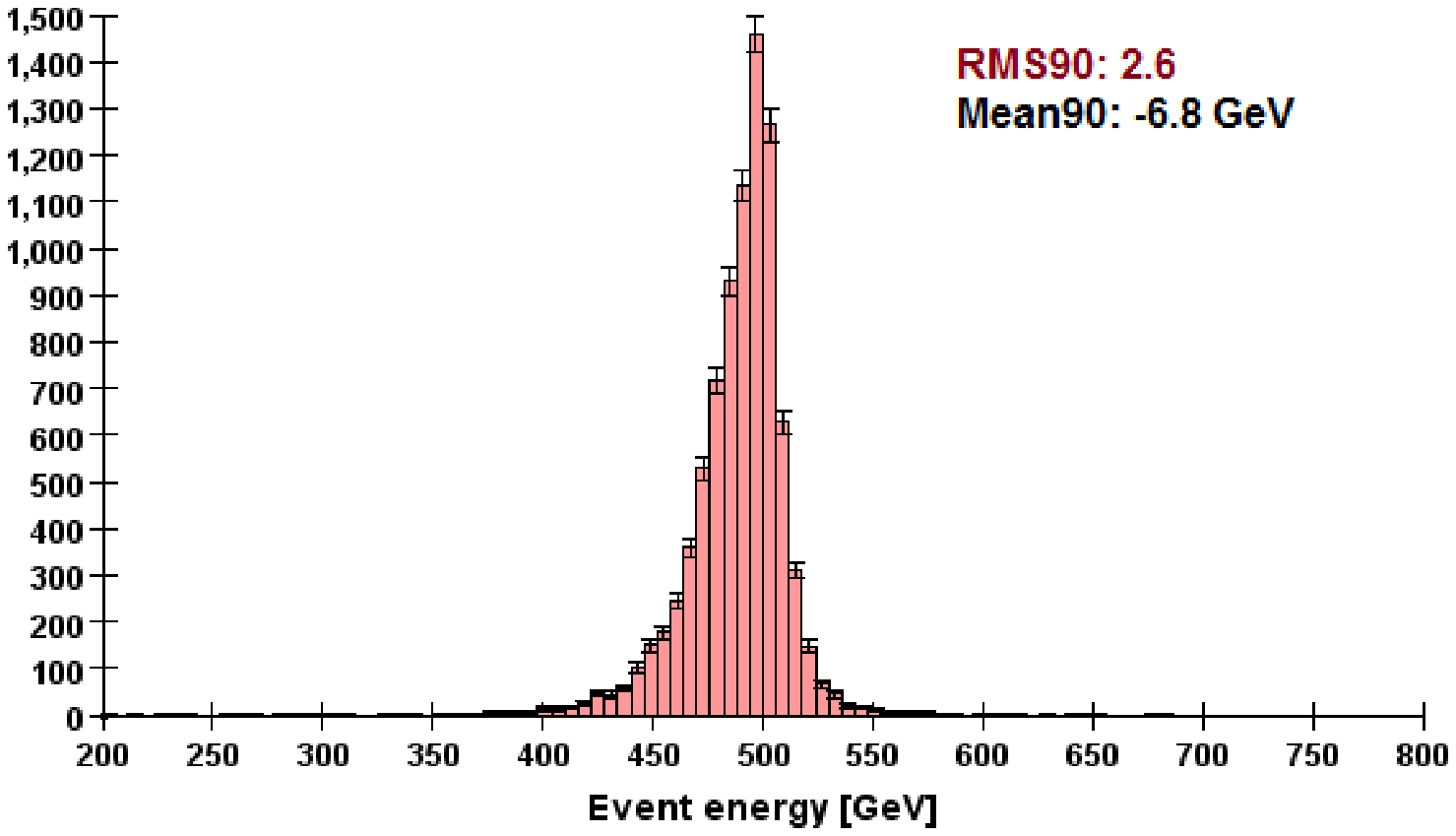}}
\caption{The energy residual distribution obtained with perfect photon finding.}
\label{fig_proceedingPlot11}
\end{minipage}
\end{figure}

\noindent
In the objective of understanding the different contributions to the resolution, MC truth information can be used in several stages of the algorithm, to replace the real reconstruction by a perfect reconstruction. We tested several cases:
\begin{itemize}
\item Perfect shower building: subclusters (MIPs, clumps and blocks) are connected to each others if the same primary particle has the dominant energy deposition in each of the subclusters. The event energy residual distribution is shown in Figure \ref{fig_proceedingPlot10}. The resolution improves from $3.1\%$ to $3.0\%$.
\item Perfect photon finding: hits in the ECal that are created by primary photons are grouped together, and the resulting clusters are identified as photons. The event energy residual is shown in Figure \ref{fig_proceedingPlot11}. The resolution in this case imroves to $2.6\%$.
\item Perfect and ideal PFA: combining together perfect photon finding and perfect shower building, gives an event energy resolution of $2.0\%$. This does not yet reflect the ideal performance of a perfect PFA: to estimate this limit, hits in the calorimeters that belong to the same primary particles are grouped together to form showers. The resolution obtained with this ideal PFA is $1.5\%$.
\end{itemize}

\section{New shower building}

A new shower building algorithm is under development. It consists from three main iterations:
\begin{itemize}
\item The aim of the first iteration is to produce a shower skeleton using tight criteria to have high purity with reasonable efficiency. The order of the tracks doesn't affect the final results since the showers are reconstructed simultaneously where the overlaps between these showers are allowed. 
\item The second iteration uses the output of the first iteration. The aim of this iteration is to increase the efficiency of the showers by adding the isolated and the ambiguous subclusters. In this iteration, the neutral showers are also reconstructed.
\item The third and final iteration uses regional and overall event energy momentum balance to achieve higher purity and efficiency.
\end{itemize}

\section{Conclusion}

The Iowa PFA is a promising algorithm for future linear colliders. Several modifications are implemented and the final energy resolution for events with 500 GeV at center of mass energy is improved from $3.5\%$ to $3.1\%$. Considering a perfect photon finding decrease this number to $2.6\%$ which shows the need to improve the photon reconstruction and identification.




\section{Bibliography}



\begin{footnotesize}


\end{footnotesize}


\end{document}